\documentclass[twocolumn]{article}
\usepackage[final]{graphics}

\font\tenmsa=msam10
\font\sevenmsa=msam7
\font\fivemsa=msam5
\font\tenmsb=msbm10
\font\sevenmsb=msbm7
\font\fivemsb=msbm5
\newfam\msafam
\newfam\msbfam
\textfont\msafam=\tenmsa  \scriptfont\msafam=\sevenmsa
\scriptscriptfont\msafam=\fivemsa
\textfont\msbfam=\tenmsb  \scriptfont\msbfam=\sevenmsb
\scriptscriptfont\msbfam=\fivemsb

\global\mathchardef\lesssim "142E

\newcommand{\slL}{\raise.15ex\hbox{$/$}\kern-.53em\hbox{$L$}}
\newcommand{\slP}{\raise.15ex\hbox{$/$}\kern-.53em\hbox{$P$}}
\newcommand{\slR}{\raise.15ex\hbox{$/$}\kern-.53em\hbox{$R$}}
\newcommand{\slQ}{\raise.15ex\hbox{$/$}\kern-.53em\hbox{$Q$}}
\newcommand{\slK}{\raise.15ex\hbox{$/$}\kern-.53em\hbox{$K$}}
\newcommand{\slSigma}{\raise.15ex\hbox{$/$}\kern-.53em\hbox{$\Sigma$}}
\newcommand{\slcalP}{\raise.15ex\hbox{$/$}\kern-.63em\hbox{$\cal P$}}

\newcommand{\be}{\begin{equation}}
\newcommand{\ee}{\end{equation}}     
\newcommand{\bea}{\begin{eqnarray}}
\newcommand{\ena}{\end{eqnarray}}

\def\build#1\over#2{\mathrel{\mathop{\kern 0pt#1}\limits_{#2}}}

\font\tenimbf=cmmib10 at 12pt
\font\sevenimbf=cmmib10 at 7pt
\font\fiveimbf=cmmib10 at 5pt
\newfam\imbf
\textfont\imbf=\tenimbf
\scriptfont\imbf=\sevenimbf
\scriptscriptfont\imbf=\fiveimbf
\def\imb{\fam\imbf\tenimbf}

\begin{document}
\date{January 20, 1999}
\title{\bf{KLN theorem, magnetic mass,\\ and thermal photon production}}
\author{
P.~Aurenche$^{(1)}$, F.~Gelis$^{(2)}$, H.~Zaraket$^{(1)}$}
\maketitle

\begin{center}
\begin{enumerate}
\item Laboratoire de Physique Th\'eorique LAPTH,\\
UMR 5108 du CNRS, associ\'ee \`a l'Universit\'e de Savoie,\\
BP110, F-74941, Annecy le Vieux Cedex, France
\item Brookhaven National Laboratory,\\
Physics Department, Nuclear Theory,\\
Upton, NY-11973, USA
\end{enumerate}
\end{center}

\begin{abstract} We study the infrared singularities associated to  ultra-soft
  transverse gluons in the calculation of photon production by a
  quark-gluon plasma. Despite the fact that the KLN theorem works in
  this context and provides cancellations of infrared singularities,
  it does not prevent the production rate of low invariant mass
  dileptons to be sensitive to the magnetic mass of gluons and
  therefore the rate to be non perturbative.
\end{abstract} 
\vskip 4mm 
\centerline{\hfill LAPTH--756/99,  BNL-NT--99/5}
\vskip 2cm

\section{Introduction}
%{\bf Position du probl\`eme:} quand on rajoute des gluons \`a la
%boucle de quarks, seule la somme des impulsions est contrainte par la
%cin\'ematique. Tous les gluons transverses sauf un peuvent donc \^etre
%arbitrairement soft, puisqu'ils n'ont pas de masse de Debye \`a
%l'ordre des boucles dures. Par comptage de puissance, chaque gluon
%suppl\'ementaire conduit \`a un facteur $g^2T/l^*$, o\`u $l^*$ est un
%cutoff introduit \`a la main pour rendre finie l'int\'egrale sur
%l'impulsion du gluon additionnel. Si ce cutoff est une masse
%magn\'etique, alors tous les ordres ont la m\^eme amplitude.

It is widely accepted that infrared singularities are generally
stronger in thermal field theories with bosons, compared to their
counterparts at zero temperature. This is due to the singular
behavior of the Bose-Einstein statistical weight at zero energy,
which affects massless bosonic fields\footnote{For a field of mass
  $m$, the statistical weight (to be evaluated on-shell in the
  real-time formalism) is bounded by $T/m$.}. As a consequence of
these stronger singularities, only partial results exist concerning
their cancellation in the calculation of observable quantities in
thermal massless theories (see \cite{KLN} for instance). So far, there
is no general translation in the language of thermal field theory of
the arguments given for this cancellation at $T=0$ by Kinoshita~\cite{Kinosh},
and  Lee and Nauenberg~\cite{LeeN1}.

The resummation of hard ther\-mal loops (HTL in the following)
\cite{HTL} cures partly this problem by giving a thermal mass to
otherwise massless fields, like gauge bosons. Nevertheless, the static
magnetic (transverse) modes remain massless in this framework and may
still generate infrared singularities, as exemplified by the
calculation of the fermion damping rate~\cite{QUARK-DAMPING}.  In QCD,
it is believed that a thermal mass for the static transverse modes is
generated non perturbatively at the scale $g^2T$, but this mass may be
too small to be an efficient regulator.

\begin{figure}[htbp]
\centerline{\includegraphics{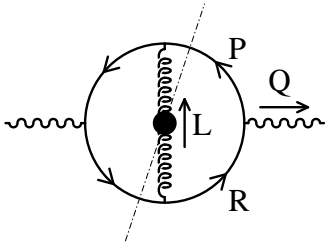}}
\caption{\em
  {2-loop dominant contribution to photon production. The blob
    indicates that the gluon propagator includes the resummation of
    hard thermal loops. The quark propagators include a thermal mass
    $M_\infty$, arising from the HTL resummation in the hard limit.}
  }
\label{fig:2-loop}
\end{figure}
A particular area where this infrared problem becomes relevant is the
thermal production of particles. In this paper, we focus mainly on the
production of photons by a quark gluon plasma. The production rates
are calculated as the imaginary part of a self-energy diagram
evaluated at finite temperature \cite{RATE}, and are expected to be
observable quantities that should come out finite in a consistent
calculation.

In a recent study \cite{BREMSS,BREMSS-1,BREMSS-2}, it has been
shown that 2-loop contributions involving the bremsstrahlung mechanism
overwhelms 1-loop contributions for the
production of a soft real photon. The insertion of an exchanged gluon in the
hard quark loop (see Fig.~\ref{fig:2-loop}) generates collinear singularities
which are power like in 2-loop diagrams while they are only
logarithmic in the 1-loop contributions: as a consequence, when these
singularities are regularized by the resummation of the thermal mass
$M_\infty\sim gT$ on the quark propagators, the two-loop diagrams get an
enhancement by powers of $g^{-1}$, where $g$ is the strong coupling constant.

The contribution of the diagram of Fig.~\ref{fig:2-loop}, although
dominated by a soft gluon, is infrared finite. In fact, even the
contribution of the transverse gluon is finite in this particular
calculation, due to kinematical constraints. Indeed, it is trivial to
see that the two delta functions corresponding to the cut quarks
$\delta(P^2-M_\infty^2)\delta((R+L)^2-M^2_\infty)$ become
$\delta(P^2-M_\infty^2)\delta(R^2-M^2_\infty)$ in the limit of
vanishing $L$, and that the latter pair of delta functions do not have
a common support if $M_\infty\not=0$: for the bremsstrahlung process
we are considering here the energies $p_0$ and $r_0$ have the same
sign and hence $P^2-M_\infty^2$ and $R^2-M^2_\infty$ cannot vanish
simultaneously, whatever the value of $Q^2$. It is therefore
kinematics, via the {\sl fermion} thermal mass, that prevents infrared
singularities in this particular topology by providing a natural
cutoff of order $gT$ on the gluon momentum $L$.  This statement was
tested in \cite{BREMSS-1} by studying the limit $M_\infty\to 0$. A
stronger divergence was found in the transverse gluon contribution,
indicating that $M_\infty$ played a role in the regularization of this
potentially dangerous contribution.

\begin{figure}[htbp]
\centerline{\includegraphics{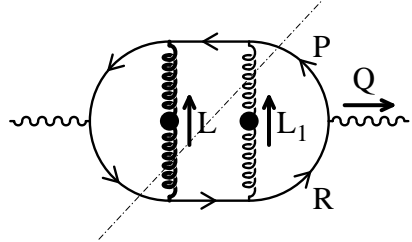}}
\caption{\em
{Example of three loop contribution for 
  photon production. One of the gluons cannot become ultra-soft due to
  kinematics, and is displayed in boldface. The other, unconstrained,
  gluon is displayed as usual.}
         }
\label{fig:3-loop}
\end{figure}

But this kinematical cut-off does not apply to additional soft gluons
one may insert in the quark loop, like in the diagram of
Fig.~\ref{fig:3-loop} for instance.
Indeed, in topologies involving more than one exchanged gluon, the
kinematical argument given above constrains only the {\sl sum} of the
momenta of the cut gluons. Therefore, we know that $L+L_1$ cannot
vanish, which tells us that (for instance) $L$ has a lower bound at
the scale $gT$, but $L_1$ can still become arbitrary
small\footnote{There is another, symmetric, contribution coming from
  the region of phase space where $L_1$ is of order $gT$ and $L$
  can become arbitrary small.} and this leads to an infrared divergence for the
cut depicted on Fig.~\ref{fig:3-loop} when the ultra-soft gluon is transverse.
%%
% Assuming a magnetic mass at some scale $\mu\sim g^2T$, it
% becomes finite and one can estimate the order of magnitude of this 3-loop
% correction. 
%%
Indeed when compared to the 2-loop diagram, the additional gluon
provides (i) two coupling constants, (ii) two quark propagators, (iii) a set of
gluon spectral density and statistical weight and (iv) the phase-space of the
additional gluon. Collecting everything, we can estimate by a crude power
counting:
\begin{eqnarray}
&&{\rm (3-loop)}\nonumber\\
&&\sim{\rm (2-loop)}\times g^2\int{d^4L_1}n_{_{B}}(L_1)
\rho(L_1)\nonumber\\
&&\qquad\qquad\times S(P+L_1)S(R+L_1)\nonumber\\
&&\sim{\rm (2-loop)}\times g^2\int_{\mu}{l_1^3 dl_1}{T\over{l_1}}
{1\over{l_1^2}} {T\over{Tl_1}}{T\over{Tl_1}}\nonumber\\
%&&\sim{\rm (2-loop)}\times g^2T\int_{\mu}{{dl_1}\over{l_1^2}}\nonumber\\
&&\sim
{\rm (2-loop)}\times {{g^2T}\over\mu}\; ,
\end{eqnarray}
where $\rho$ is the spectral function of the additional gluon, where
$\mu$ is introduced as a regulator on the integral over $l_1$.  We
used the fact that the quarks are hard, and mostly
on-shell\footnote{The role played by the small off-shellness of the
  additional quarks will be considered later on in this paper.}
because of the cut crossing the quark loop. It is important to stress
here that each fermion propagator brings an extra factor $l_1$ in the
denominator thus contributing to the infrared sensitivity of the above
expression. The conclusions of this naive power counting are the
following:
\begin{itemize}
\item If the additional gluon is longitudinal, its cutoff is a
  thermal mass of order $gT$ , and
  the corresponding contribution is suppressed by one power of $g$
  compared to the 2-loop one.
\item If the additional gluon is transverse it is natural to assume the
  regulator to be the magnetic mass $\mu\sim g^2T$, and we have ${\rm
  (3-loop)}\sim{\rm (2-loop)}$.
\end{itemize}

%The infrared problem illustrated here on the 2-point function occurs also
%for n-point functions with massless external lines: attaching ultra-soft gluons
%to a cut internal line will likewise generate infrared divergences.

Therefore, it seems that if we keep adding transverse gluons in the quark loop,
we generate contributions that are all of the same order of magnitude. This
fact is very similar to the argument given by Linde for the breakdown of
perturbation theory in thermal QCD, although in the different context of the
calculation of the free energy \cite{Linde1}.

%Cette estimation ne prend pas en compte d'\'eventuelles compensations
%entre coupures, qui seraient une g\'en\'eralisation thermique du
%th\'eor\`eme KLN. Il convient donc d'\'etudier la possibilit\'e de
%telles compensations, et de comparer leur efficacit\'e avec la masse
%magn\'etique en tant que r\'egulateur.

There is nevertheless one reason why this power counting may be too
naive.  One should indeed keep in mind that this estimate is valid
only for a given cut through the 3-loop diagram. It does not take into
account potential compensations that may occur when one is summing all
the possible cuts. In this paper, we are going to study in more detail
this possibility, and its interplay with a magnetic mass at the scale
$g^2T$.

%We show in section \ref{sec:kln-general} that there are cancellations
%between the different cuts one can draw through higher order diagrams,
%and that these compensations are enough to make the corresponding
%contributions finite without the need of a magnetic mass. We have
%therefore two mechanisms of regularization for the transverse IR (the
%magnetic mass, and the cancellations a la KLN) singularities, and we
%compare their efficiency in section \ref{sec:competition}. It appears
%that the magnetic mass is the relevant regulator when one is
%considering the production of soft quasi-real photons, while one can
%safely ignore it for hard, or virtual enough, photons.

\section{Infrared cancellations}
\label{sec:kln-general}
%Dans cette section, on donne un argument assez g\'en\'eral en faveur
%d'une g\'en\'eralisation thermique du th\'eor\`eme KLN, valable
%topologie par topologie, et consistant en une compensation entre les
%diverses coupures au sein d'un diagramme donn\'e.
An important feature of the above example is the fact that the {\sl
  quark propagators} participate in the overall infrared divergence of
the diagram. In fact, if some quark propagators were not becoming
singular in the IR limit, the diagram would have been finite by power
counting.  This can be generalized to a topology with an arbitrary
number of exchanged gluons (but without 3- and 4-gluon
vertices)\footnote{In this paper, we are considering only abelian
  topologies, since this is enough for our purpose of studying the
  interplay between the magnetic mass and possible cancellations. Later on,
  we indicate why the arguments given here cannot be applied to
  non-abelian topologies.}.  Indeed, for these topologies, the number
of loops $L$ is related to the number of gluons $n_g$ by:
\begin{equation} 
L=1+n_g\; . 
\end{equation}
One of the $L$ loop integrals is an integration $d^4P$ over the quark
momentum which is hard, and is not concerned by the IR problem. The
remaining $L-1$ integrals are over the momenta of the $n_g$ soft
gluons. The fact that $L-1=n_g$ tells that even if each gluon comes
with the singular factor $n_{_{B}}(l^o)\rho(L)$, it is accompanied by
a phase space $d^4L$ which is enough to make the integral finite.  For
these topologies, it is the quark propagators which are ultimately
responsible for the IR divergences. Indeed, it is trivial to see that
if one quark propagator is cut, then some other quark propagators
become infinite when the gluon momenta go to zero. This is the reason
why we are going to focus on the quarks, and do not care about the
gluon propagators.

\begin{figure}[htbp]
\centerline{\includegraphics{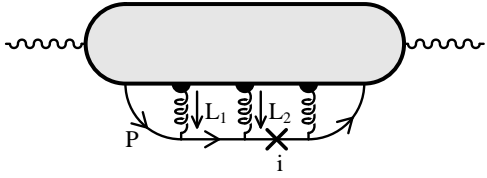}}
\caption{\em
{Generic configuration for which 
  cancellations in the infrared sector occur. Again, the gluons all
  include the HTL correction. The gluons do not need to be parallel for
  the argument to work, since the gray area hides the details about
  how the gluons are attached to the upper quark line. The index $i$,
  running from $0$ to $n$, indicates which quark propagator is cut on
  the lower fermion line.}
        }
\label{fig:kln-generic}
\end{figure}

It is rather easy to show that there are cancellations at the level of
the quark propagators occuring in the calculation of the imaginary
part of $\Pi^\mu_\mu$. Let us illustrate that in the reasonably
general situation where we have a single quark loop to which are
connected the two external photons, and an arbitrary number of
internal gluons (but without non-abelian couplings, see
Fig.~\ref{fig:kln-generic}).
%{\em Sans quasiment rien changer a la suite on pourrait ajouter autant de
%lignes externes que l'on veut sur le ``blob" de la figure et on aurait des
%considerations sur les fonctions de Green a N-points. Pour avoir une
%compensation complete des divergences sur la ligne de quark il suffit de
%considerer toutes les coupures entre 2 lignes externes. On pourrait aussi dire
%quelque part que si on avait une ligne de gluons nos compensations se feraient
%de facon similaire.}
To simplify, we detail only the lower fermion line, and hide the
details of the upper line in a complicated function we do not need to
specify.

Since we are going to demonstrate cancellations among the various
contributions to the dispersive part of the photon polarization
tensor, it is convenient to work in the R/A formalism \cite{RAform},
in which cutting rules exist that are both very simple and very close
to the $T=0$ ones \cite{RAcut}\footnote{If the R/A amplitudes are
  finite, then so do the time-ordered ones. Nevertheless, checking the
  compensations for an arbitrary number of loops with the rules found
  in \cite{CTPcut} would be awkward since it is not possible to write
  all the contributions as cut diagrams. If one insists on using the
  CTP formalism as an exercise, then the rules given in \cite{CTPcut1}
  are a better starting point.}.  Therefore, all the contributions to
the imaginary part of $\Pi_{_{R}}$ can be obtained by cuts dividing
the diagram into two connected pieces, each of these parts containing
at least one external leg.
%{\em Nous n'avons jamais
%  explique que cette regle de sommer sur les coupures est valable dans
%  le formalisme R/A (dans le formalisme 1/2 il faudrait faire appel
%  d'urgence a Niegawa pour rassembler toutes les sommes sur les
%  indices 1 et 2 et en tirer des conclusions). je crois qu'il faut
%  dire qqes mots sur les regles a suivre en R/A}.

%In order to simplify the discussion and to exhibit the smallest set of
%terms needed to have a cancellation, it is possible to keep only the
%positive energy part of the quark propagators.  
%%{\em Le fait de ne
%%  garder que l'energie positive est completement justifiee si la
%%  boucle consideree est dure. Les termes negliges sont suprrimes par
%%  des puissances de T.}  
%By doing so, we exclude from our
%considerations the processes in which the quark becomes (or
%annihilates with) an antiquark\footnote{That means we are considering
%  the emission of a photon by a quark (or antiquark) undergoing
%  multiple scatterings in the medium, as well as some loop corrections
%  to this process.}. 

By summing over the index of the cut quark on the lower fermion line,
the contribution of the diagram of Fig.~\ref{fig:kln-generic} can be
written as\footnote{In the R/A formalism, we pick the most singular
  piece for each gluon, i.e.  $n_{_{B}}(l^0_i)\rho(L_i)$.  Failing to
  do this, we would get a contribution suppressed compared to the
  2-loop result.
%{\em This amounts to considering only cut gluons
%    leading to the most singular behavior.}
}
 % {\em On pourrait cacher n-F(r)-n-F(p) dans la function F(P, L-i) sans perdre
%  en generalite et ca nous epargnerait des explications. Dans ce cas changer
%  aussi eq. (4)}
\begin{eqnarray}
&&\!\!\!\!\!\!\!\!\!\!\!\!A\equiv\int\!{d^4P}
\Big[\!\prod_{i=1}^{n}{d^4L_i}n_{_{B}}(l^0_i)\rho(L_i)\;{\rm Trace}\Big]
F(p^0,p,\{L_i\})\nonumber\\
&&\!\!\!\!\!\!\!\times\sum_{i=0}^{n}
\delta((P+K_i)^2-M^2_\infty)\prod_{j\not=i}{1\over{(P+K_j)^2-M^2_\infty}}
\end{eqnarray}
where the function $F$ hides all the details about the denominators on
the upper quark line\footnote{This function does not depend on the
  position of the cut on the lower quark line, but depends on the
  position of the cut on the upper line.} (as well as the fermionic
statistical factors, which do not play any role in the following), and
where we have defined
\begin{equation}
K_i\equiv\sum_{a=1}^{i}L_a
%\; ,\;
% \omega_i\equiv \sqrt{({\imb p} + {\imb k}_i)^2+M^2_\infty}
\; .
\end{equation}
We can now use the $\delta$ functions to perform explicitly the
integral over $p^0$, which gives after splitting the propagators into
positive and negative energy terms
\begin{eqnarray}
&&\!\!\!\!\!\!\!\!A=\int{d^3{\imb p}}\Big[\prod_{i=1}^{n}{d^4L_i}
n_{_{B}}(l^0_i)\rho(L_i)\;{\rm Trace}\Big]\prod_{j=0}^n{1\over{2\omega_j}}
\nonumber\\
&&\!\!\!\!\!\!\times
\sum_{\build{i=0}\over{\epsilon=\pm 1}}^{n}
\!F(\alpha_i^\epsilon,p,\{L_i\})
\prod_{j\not=i}\!\Big[{1\over{\alpha_i^\epsilon-\alpha_j^+}}
-{1\over{\alpha_i^\epsilon-\alpha_j^-}}\Big]
\label{eq:kln-1}
\end{eqnarray}
where we denote $\omega_i\equiv \surd({({\imb p} + {\imb
    k}_i)^2+M^2_\infty})$ and
$\alpha_j^\epsilon\equiv-k_j^0+\epsilon\omega_j$.  According to this
definition, all the $\alpha_i^\epsilon$ become equal to
$\epsilon\surd({\imb p}^2+M^2_\infty)$ when the gluon momenta go to
zero. We see that denominators where both $\alpha_i$'s carry the same
sign vanish in this limit.  The IR singularities therefore show up
in the vanishing denominators $\alpha_i^\pm-\alpha_j^\pm$.
%{\em On pourrait rajouter un ptit commentaire intuitif du genre:
%It is obvious at this point that some compensations occur in the above equation
%since for every denominator $\alpha_i-\alpha_j$ with a numerator 
%$F(\alpha_i,p,\{L_i\})$ appears a denominator $\alpha_j-\alpha_i$ with
%a numerator $F(\alpha_j,p,\{L_i\})$. We now prove rigourously that, indeed,
%these denominators appear in ...............}
Only the second line in Eq.~(\ref{eq:kln-1}) is relevant in the
following discussion, and it can be compactly rewritten as
\begin{equation}
\sum_{\build{i=0}\over{\{\epsilon_i=\pm_1\}}}^n
\!F(\alpha_i^{\epsilon_i},p,\{L_i\})
\prod_{j\not=i}{{\epsilon_j}\over
{\alpha_i^{\epsilon_i}-\alpha_j^{\epsilon_j}}}\; .
\end{equation}
One can simply observe that for every denominator
$\alpha_i^\pm-\alpha_j^\pm$ with numerator $F(\alpha_i^\pm,p,\{L_i\})$
appears a denominator $\alpha_j^\pm-\alpha_i^\pm$ with a numerator
$F(\alpha_j^\pm,p,\{L_i\})$ (all the other denominators being the
same). The simple poles therefore cancel trivially. This can be
extended to the more complicated situation where more than two
$\alpha_i$s tend to a common value, which amounts to prove that these
denominators appear in a combination that remains finite for any
configuration of the $\alpha_i$s. For that purpose, let us consider an
expression like\footnote{At this stage, we can drop all the
  superscripts $\epsilon_i$ since the compensations occur in fact for
  each given set of $\epsilon_i$s.}
\begin{equation}
F_n\equiv\sum_{i=0}^{n}F(\alpha_i)
\prod_{j\not=i}{1\over{\alpha_i-\alpha_j}}
\end{equation}
and show that such a quantity is always finite provided some
regularity property of the function $F$. The shortest way to see that
is to notice that $F_n$ is the leading coefficient of the Lagrange
polynomial of degree $n$ that interpolates between the points
$(\alpha_i,F(\alpha_i))$:
%{\em Ne pourrait-on pas ecrire explicitement le polynome de Lagrange en
%question ce qui permettrait aux beotiens comme moi d'etre moins impressionne
%par ce nom prestigieux et aussi leur eviterait d'avoir a reflechir trop
%longtemps sur la construction explicite du dit polynome!}
\begin{equation}
P_n(x)\equiv\sum_{i=0}^nF(\alpha_i)\prod_{j\not=i}
{{x-\alpha_j}\over{\alpha_i-\alpha_j}}=F_n\,x^n+\cdots
\end{equation}
As such, $F_n$ is finite for every values of the $\alpha_i$'s if the
function $F$ is $n$ times differentiable.  Indeed, if several points
$(\alpha_i,F(\alpha_i))$ collapse into a single point of multiplicity
$m$, the Lagrange polynomial has a finite limit, and coincides with
$F$ and its first $m-1$ derivatives at this point. One can note that
this argument also applies to insertions of self-energies on the quark
line. Indeed, the only peculiarity of self-energy insertions is that
they force several $\alpha_i$s to have the same value even if the
$L_i$ do not tend to zero; and the above proof works for any
configuration of the $\alpha_i$s, whatever the cause that makes them
equal is.

Although the above argument has been presented in the case of a
2-point function, it still works for the dispersive part of any
$n$-point function (just attach more photons or gluons to the shaded
part of Fig.~\ref{fig:kln-generic}).

\section{Competition with the magnetic mass}
\label{sec:competition}
%Ensuite, on peut se sp\'ecialiser \`a une topologie particuli\`ere \`a
%trois boucles (la topologie en \'echelle pour rester simple), et
%montrer explicitement comment fonctionne cette compensation, et
%comment elle interagit avec la pr\'esence d'un masse magn\'etique. Il
%s'agit de comparer le cutoff fourni par cette compensation avec la
%masse magn\'etique. On peut terminer la section par un diagramme du
%plan $(Q^2,q_0)$ montrant dans quel domaine on a une sensibilit\'e vis
%\`a vis de la masse magn\'etique (et \'egalement une contribution \`a
%l'ordre dominant de tous les diagrammes), et dans quel domaine au
%contraire l'approche perturbative fonctionne et le r\'esultat est
%domin\'e par la contribution \`a deux boucles.

The result of the previous section is that the product of all the quark
propagators remains finite in the IR sector when the sum over the cuts has been
performed. When considering the imaginary part it shows  that, for a fixed cut
on the upper line, there is no singularity in the propagators of the lower
line. The argument should be repeated for the upper line, which is made finite
by summing over all the ways of cutting it. These cancellations between
different cuts occur within a given topology, and correspond to compensation
between real and virtual corrections. They can therefore be seen as a form of
the KLN theorem.

We point out some differences with the usual version of the KLN
theorem at zero temperature where the exchanged gluons are bare
gluons. At finite temperature we deal with resummed gluons and the
most dangerous divergences arise when cutting space-like transverse
gluons which are shielded by a ``small'' magnetic mass. The time-like
gluons are shielded with a ``large'' thermal mass of order $gT$ and do
not lead to any infra-red problems in the context of this study.

\begin{figure}[htbp]
\centerline{\resizebox*{7.5cm}{!}{\includegraphics{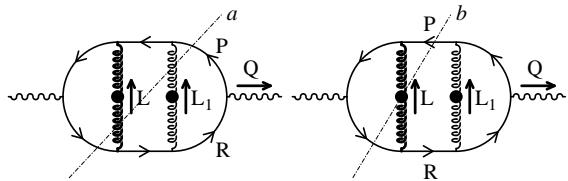}}}
\caption{\em
{Notations used for the 3-loop contributions.}
\label{fig:3-loop-cuts}
        }
\end{figure}

We are now going to apply the above considerations to study explicitly
the 3-loop example already presented in Fig.~\ref{fig:3-loop}. As we
have already seen in the introduction, one of the gluon momenta has a
lower bound thanks to kinematics, while the other gluon is not
constrained in the ultra-soft limit.  Let us choose the gluon on the
right (momentum $L_1$) to be ultra-soft while momentum $L$ remains
soft.  This kinematical constraint prevents the propagators of the
lower quarks to become infinite when $L_1\to 0$: they always remain
off-shell by some amount controlled by $M_{\infty}$.  The constraint
also simplifies the pattern of cancellation of infrared divergences in
the upper line since the propagators of momentum $P+L+L_1\approx P+L$
when $ L_1 \to 0$ cannot diverge for the same reason. In consequence we
need only the two cuts depicted on Fig.~\ref{fig:3-loop-cuts} to get
rid of all the zeroes in the denominators\footnote{We also have to
  take into account the other possibility with $L_1$ soft and $L$
  ultra-soft which is easily done by multiplying the final result by a
  factor $2$.}.
It turns out to be convenient to perform the change of variable
$P+L_1\to P$ in the second contribution, as indicated on
Fig.~\ref{fig:3-loop-cuts}. Moreover, we neglect $L_1$ in front of $L$
whenever these two impulsions appear together (like in $L_1+L\approx
L$). With these notations and approximations, one can readily see
that
the additional ultra-soft gluon brings the following factors, to be
multiplied by the 2-loop integrand (i.e. the integrand for the diagram
of Fig.~\ref{fig:2-loop}), respectively for the cuts $(a)$ and $(b)$:
\begin{eqnarray}
&&F_{a,b}=g^2\int{{d^4L_1}\over{(2\pi)^4}}
(2P_\sigma)(2R_\rho)P^{\sigma\rho}_{_{T}}(L_1)
\nonumber\\
&&\ \ \times
{{\rho_{_{T}}(l_1^0,l_1)n_{_{B}}(l^0_1)}
\over{((P\pm L_1)^2-M_\infty^2)((R\pm L_1)^2-M_\infty^2)}}
%\nonumber\\
%&&F_b=g^2\int{{d^4L_1}\over{(2\pi)^4}}
%{{(2P_\sigma)(2P_\rho)P^{\sigma\rho}_{_{T}}(L_1)
%\rho_{_{T}}(l_1^0,l_1)n_{_{B}}(l^0_1)}
%\over{(P-L_1)^2(R-L_1)^2}}
\; ,
\end{eqnarray}
where $P^{\sigma\rho}_{_{T}}$ is the transverse projector, and
$\rho_{_{T}}$ is the transverse spectral function of the gluon (only
the transverse mode of the ultra-soft gluon is relevant, the other
gluon can be transverse or longitudinal, as in the 2-loop
calculation). The factors of form  
$P_\sigma/((P\pm L_1)^2-M_\infty^2)$ come from the fermion propagators in
the hard momentum approximation.
We can further simplify these factors by evaluating
them at $P^2=M_\infty^2$ since they are to be multiplied by the 2-loop
integrand which contains a $\delta(P^2-M^2_\infty)$. Noticing that
(see eq. (46) in \cite{BREMSS-1})
\begin{equation}
P_\sigma R_\rho P^{\sigma\rho}_{_{T}}(L_1)\approx -pr(1-\cos^2\theta_1)\; ,
\end{equation}
where $\theta_1$ is the angle between the vectors ${\imb p}$ and
${\imb l}_1$, it is easy to perform the angular integral explicitly,
which gives:
\begin{eqnarray}
&&\!\!\!\!\int d\Omega_1 {{1-\cos^2\theta_1}
\over{((P\pm L_1)^2-M_\infty^2)((R\pm L_1)^2-M_\infty^2)}}\nonumber\\
&&\quad\approx
{\pi\over{prl_1^2}}\Big( \Big[x \!\pm{\Delta\over{4rl_1}}\Big]
\ln\Big|{{2rl_1(x+1)\pm\Delta}\over{2rl_1(x-1)\pm\Delta}}\Big| -\!1 \Big) \nonumber\\
&&\quad
\pm{{\pi(1-x^2)}\over{pl_1\Delta}}\Big[\ln\Big|{{x+1}\over{x-1}}\Big|\nonumber\\
&&\qquad\qquad\quad
-\ln\Big|{{2rl_1(x+1)\pm\Delta}\over{2rl_1(x-1)\pm\Delta}}\Big|\Big]\; ,
\end{eqnarray}
where we denote $x\equiv l_1^0/l_1$ and $\Delta\equiv 2P\cdot Q+Q^2$.
The general result established in the previous section says that the
sum over the cuts for this quantity should have a finite limit when
$l_1\to 0$ at fixed $x$ (i.e. when the 4 components of $L_1$ tend to
zero). One can see that this is not the case on the
contributions of the individual cuts, since we have:
\begin{eqnarray}
&&\!\!\!\!\!\!\!\!\int d\Omega_1 {{1-\cos^2\theta_1}
\over{((P\pm L_1)^2-M_\infty^2)((R\pm L_1)^2-M_\infty^2)}}
\nonumber\\
&&\build{\approx}\over{l_1\to 0}\pm{{\pi(1-x^2)}\over{pl_1\Delta}}
\ln\Big|{{x+1}\over{x-1}}\Big|\pm{{4\pi x}\over{pl_1\Delta}}\; ,
\end{eqnarray}
(where we have neglected terms regular in $l_1$) but occurs trivially
for the sum of the two cuts since the singular ($l_1^{-1}$) part drops
out. The interesting quantity is in fact the scale at which this
cancellation occurs. The only dimensional quantity to which $l_1$ can
be compared with in the above expressions is $l_1^*\sim \Delta/r$.
Carrying out the integration on $x, l_1$ we find the following finite
result:
\begin{equation}
F_{a+b} \approx {{g^2Tr}\over{2\Delta}}\; .
\label{muequ0}
\end{equation}

%Comme on a besoin ici de l'expression de $M^2_{eff}$ en fonction de
%$Q^2$ et de $q_0$, on peut donner l'expression de cette masse
%effective dans le cas le plus g\'en\'eral (photons pouvant \^etre
%virtuels durs, et avec une largeur pour les quarks). Cela permet de
%baliser le terrain au sujet des choses qu'on a \`a dire sur les
%photons virtuels durs et de l'effet de la largeur, tout en se donnant
%du temps pour comprendre les complications associ\'ees au processus
%d'annihilation.

When we add a magnetic mass $\mu$ into the game, we have two
regularization mechanisms in competition, and we must compare the
scales at which they operate, the most efficient being the one that
has the largest scale. A convenient way to introduce\footnote{This
  does not tell much about the way the magnetic mass enters in the
  gluon propagator: it just tells that the self-energy of the
  transverse gluon satisfies $\Pi_{_{T}}(x=0)=\mu^2$, which is a
  definition analogous to that of the Debye screening mass.} the
magnetic mass is via the following sum rule
\begin{equation}
  \int\limits_{-\infty}^{+\infty}{{dx}\over{2\pi}}
  {{\rho_{_{T}}(l_1,x)}\over{x}}B(x)\approx
  {{B(0)}\over{l^2_1+\mu^2}}\; , 
\end{equation} 
which is a reasonable approximation if $B$ is not singular at $x=0$ and
does not increase too much when $x$ becomes large.  Adding the two cuts and
approximating $n_{_{B}}(l^0_1)\approx T/l_1x$, we find that the 2-loop
integrand is to be multiplied by the factor 
\begin{eqnarray}
  &&F_{a+b}\approx {{g^2Tr}\over{\pi^2\Delta}}\int\limits_{0}^{+\infty}
  {{l_1\,dl_1}\over{l_1^2+\mu^2}}
  \Big\{\ln\Big|{{\Delta+2rl_1}\over{\Delta-2rl_1}}\Big|
  \nonumber\\
&&\qquad-{\Delta\over{rl_1}}\Big[{\Delta\over{4rl_1}}
  \ln\Big|{{\Delta+2rl_1}\over{\Delta-2rl_1}}\Big|-1\Big] \Big\}\; .
\end{eqnarray}
We can now give analytical limits for this factor in two cases. If the
magnetic mass dominates over the scale $l_1^*$ defined
above, we have
\begin{equation}
F_{a+b}\build{\approx}\over{r\mu\gg\Delta} {{g^2T}\over{\pi\mu}}\sim 1\; ,
\end{equation}
while in the opposite limit ($l_1^* \ll \mu$) we recover
Eq.~(\ref{muequ0}).
\begin{equation}
F_{a+b}\build{\approx}\over{r\mu\ll\Delta} {{g^2Tr}\over{2\Delta}}\ll 1\; .
\end{equation}
Let us mention that if the additional gluon in Fig.~\ref{fig:3-loop-cuts} is
longitudinal a similar calculation as above goes through where $\mu$ is to
be replaced by the Debye mass of order $gT$, so that the factor $F_{a+b}$ is
of order $g\ll 1$.

%%%\begin{equation}
%%%F_{a+b}\build{\approx}\over{p\mu\ll\Delta} {{g^2Tp}\over{\pi\Delta}}
%%%({{\pi^2}\over{2}}-1)\ll 1\; .
%%%\end{equation}

The conclusion is therefore the following: 
\begin{itemize}
\item If the magnetic mass $\mu$ is the relevant regulator, then the
  3-loop diagram gives a ``correction" of order one to the prefactor
  of the 2-loop result, and this is likely to be the case for higher
  loop corrections also. In this regime, the photon production rate is
  sensitive to the magnetic mass, and one must resum an infinite
  series of diagrams. 
%%%\cite{WorkIP1}.
\item If on the contrary, the scale $\Delta/r$ dominates or if the
  regulator on the gluon propagator is the Debye mass, then the 3-loop
  diagrams lead to a negligeable correction to the 2-loop result, and
  it is expected that this hierarchy will be valid between 4-loop and
  3-loop,...
%%%  
%%%  je supprimerais ce paragraphe car il est repris + en detail plus bas
%%%
%%%  In this regime, the photon production rate is already
%%%  under control at two loops. This is true as long as the dynamics of
%%%  photon production is dominated by soft-gluon exchanges which is the
%%%  case when the cut-off $\Delta$ satisfies $\Delta \ll q_0 T$
  \cite{BREMSS-1}.
\end{itemize}

\begin{figure}[htbp]
\centerline{\resizebox*{!}{5cm}{\rotatebox{-90}{\includegraphics{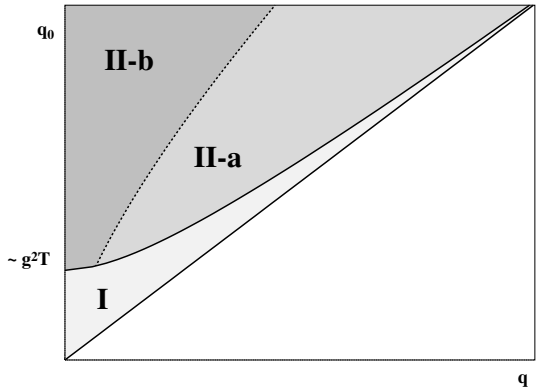}}}}
\caption{\em
  {Comparison of $\mu$ and $\Delta/r$ in the $(q,q_0)$ plane. In
    region I, the magnetic mass is the most important regulator, the
    3-loop contribution is as large as the 2-loop one. In regions II-a
    and II-b, the magnetic mass is a subdominant regulator, and the
    higher-loop contributions are subdominant. In region II-b, $L$ can
    become hard and some of our approximations become invalid (see
    \cite{BREMSS-2}).}  }
\label{fig:q-plane}
\end{figure}

It remains now to make a bit more explicit the comparison between
$\Delta/r$ and $\mu$. For that purpose, let us recall that the
expression of $\Delta=2P\cdot Q+Q^2$ can be taken from the 2-loop
calculation\cite{BREMSS}:
\begin{equation}
\Delta\approx pq_0\Big[1-\cos\theta+{{M^2_{\rm eff}}\over{2p^2}}\Big]\; ,
\end{equation}
where $\theta$ is the angle between ${\imb p}$ and ${\imb q}$, and
where\footnote{The formula for $M^2_{\rm eff}$ has been extended here
  to hold for hard slightly virtual photons \cite{ElseW1}.}
\begin{equation}
M_{\rm eff}^2=M^2_\infty+{{Q^2}\over{q_0^2}}pr\; .
\label{eq:meff}
\end{equation}
Due to the extremely singular nature of the integral over $\theta$
\cite{BREMSS,BREMSS-1}, the order of magnitude of $\Delta$ is $q_0
M^2_{\rm eff}/p$. Taking into account the fact that $r\approx p+q_0$
and $p\sim T$, we divide the $(q,q_0)$ plane in two regions where
respectively $\mu$ or $\Delta/r$ dominates (see
Fig.~\ref{fig:q-plane}).
In region I, {\em i.e.} roughly the region of small virtuality, the
magnetic mass is the most important regulator and, as seen above, the
3-loop contribution is as large as the 2-loop contribution: the
production mechanism becomes non-perturbative and resummations should
be considered. Increasing the virtuality $Q^2$ of the photon increases
$\Delta$ (see Eq.~(\ref{eq:meff})) which eventually becomes the
dominant cut-off and one enters region II where the infrared
sensitivity of 3-loop (and higher loop diagrams) becomes subdominant.

In region II-b, the production mechanism receives a contribution from
large gluon momentum and the approximations done in the previous
calculations may become incorrect. However, it was found by an
explicit calculation \cite{BREMSS-2} that the expressions valid in
region II-a could be safely extrapolated to the case of soft virtual
photon at rest ($q=0$) in region II-b.
However, there is the possibility that 3-loop diagrams give important
contributions in region III due to hard gluon exchanges, but this is a
different story.

\section{Summary and outlook}
%Il existe une r\'egion (photons pas trop durs, et pas trop virtuels)
%o\`u on a une sensibilit\'e vis \`a vis d'une masse magn\'etique. Une
%\'etude sp\'ecifique et des resommations au del\`a des boucles dures
%s'imposent dans cette r\'egion. Dans les autres r\'egions, le calcul
%\`a deux boucles est suffisant, mais doit \^etre compl\'et\'e pour les
%photons durs virtuels, et pour inclure l'effet d'une largeur.
In this paper, we have considered higher order corrections to photon
production by a quark-gluon plasma, with emphasis on the infrared
singularities due to ultra-soft transverse gluons. When one sums over
all the possible cuts through a given abelian topology, there are
cancellations that prevent the quark propagators from becoming
infinite, making the higher-loop corrections infrared finite.

The generalization of this result to non-abelian topologies is not
straightforward. Indeed, the above argument is flawed already at the
stage of counting the number of gluon propagators. Loosely speaking,
because of the 3- and 4-gluon couplings, the number of gluons can be
larger than the number of independent momenta. The loop counting for
QCD gives for the number of gluon integrations
\begin{equation}
L-1=n_g-(n_3+n_4)\le n_g\; ,
\end{equation}
where $n_3$ and $n_4$ are respectively the number of 3- and 4-gluon
vertices. As a consequence, unlike in the abelian topologies for which
$L-1=n_g$, the gluon propagators play an active role in the KLN
compensations (if any), and much more elaborate arguments are
required.  As a side note, it is also impossible to apply in the
thermal case the heuristic power counting argument of Poggio and Quinn
\cite{Poggio}, since some gluons have an infrared count of $-3$
(instead of $-2$ at $T=0$) because of the Bose-Einstein statistical
weight.

We have also considered the possible competition of this mechanism of
regularization with a hypothetical magnetic mass of order $g^2T$. The
result is that there exists a region in the $(q,q_0)$ plane (roughly
speaking, for photons of small invariant mass) where the magnetic mass
is the dominant regulator, and where the higher-loop corrections are
as large as the two-loop result, and should be resummed.

%%%
%%% {\bf In the case one considers only the multiple exchange of gluons
%%%  shielded by the Debye mass (longitudinal gluons) we find that the
%%%  production rate of photons is perturbative in the sense that higher
%%%  loop diagrams are suppressed by powers of the coupling constant. In
%%%  that respect we disagree with a study by Cleymans, Redlich and
%%%  Goloviznin who found that multiple exchanges of Debye screened
%%%  gluons lead to a Landau-Pomeranchuck type suppression of soft photon
%%%  production in a quark-gluon plasma. This result is based on a study
%%%  of photon radiation by ultra-asymptotic quarks (of energy much
%%%  larger than the temperature of the plasma) and then extrapolating
%%%  the results to the case where the quarks have an energy of the order
%%%  of the temperature. These differences in the initial kinematical
%%%  assumptions may explained the difference in the results.}
%%%

Further extensions of this work include the treatment of non-abelian
topologies, the study of resummations in the region where the magnetic
mass dominates, the extension of the 2-loop calculation for hard
massive photons, and also the study of how the finite lifetime of
quarks in a plasma affects the collinear singularities found at
2-loop.

\section{Acknowledgements}
We thank R.D. Pisarski, F. Guerin and E. Braaten for useful
discussions and comments. F.G. was supported by DOE under grant
DE-AC02-98CH10886.

\end{document}